%% file: air.tex
\documentclass[sigconf]{acmart}
\renewcommand\footnotetextcopyrightpermission[1]{} 
\usepackage{booktabs} 
\usepackage{amsmath,amsfonts,amssymb,epsfig,epstopdf,url,array}

\usepackage{amsthm}
\usepackage{color,soul}
\usepackage{algorithmic}
\usepackage{balance}
\usepackage[ruled]{algorithm2e}
\usepackage{stmaryrd}
\usepackage{centernot}
\usepackage{tikz}
\usepackage{booktabs}
\usepackage{pdfpages}
\usepackage{graphicx}
\usepackage{subfigure}
\usepackage[font=small,skip=8pt]{caption}
\usepackage{tabularx}
\usepackage{comment}
\usepackage{multirow}
\usepackage{multicol}  
\usepackage{bigstrut}
\usepackage{pdflscape}
\usepackage{rotating}
\usepackage{float}
\usepackage{lipsum}

\usepackage{enumitem}
\usetikzlibrary{plotmarks,shapes,arrows,chains,hobby,backgrounds,calc,trees}
\usepackage{makecell}
\usepackage{dblfloatfix}    
\usepackage{pdfcomment} 
\usepackage{etoolbox}
\usepackage{colortbl}
\usepackage{balance}

\definecolor{lightgray}{gray}{0.9}
\theoremstyle{definition}

\theoremstyle{definition}

\theoremstyle{definition}

\begin{document}



\title{Engineering Human Values in Software \\ through Value Programming}


\author{Davoud Mougouei}
\affiliation{%
	\institution{School of Computing and IT, University of Wollongong, Australia}
}
\email{dmougouei@gmail.com}
%

\renewcommand{\shortauthors}{}
\renewcommand{\shorttitle}{}
	

\begin{abstract}
\input{abstract}

\end{abstract}

%
%
%


\maketitle
\input{introduction}
\input{valueprogramming}

\input{proposed}

\input{conclusion}

\bibliographystyle{ACM-Reference-Format}
\balance
\bibliography{ref} 

\end{document}

%% file: abstract.tex
Ignoring human values in software development may disadvantage users by breaching their values and introducing biases in software. This can be mitigated by informing developers about the value implications of their choices and taking initiatives to account for human values in software. To this end, we propose the notion of \textit{Value Programming} with three principles: \textbf{(P1)} annotating source code and related artifacts with respect to values; \textbf{(P2)} inspecting source code to detect conditions that lead to biases and value breaches in software, i.e., \textit{Value Smells}; and \textbf{(P3)} making recommendations to mitigate biases and value breaches. To facilitate value programming, we propose a framework that allows for automated annotation of software code with respect to human values. The proposed framework lays a solid foundation for inspecting human values in code and making recommendations to overcome biases and value breaches in software. 

%% file: introduction.tex
\section{Introduction and State of the Art}
\label{sec_introduction}

Software products are manufactored and used by humans, and therefore need to account for their values~\cite{perera2019towards,mougouei2018operationalizing,hussain2018integrating,perera2019study}. A prominent model of human values is proposed by Schwartz~\cite{schwartz1987toward,lee2019testing} (Table~\ref{values}) in social sciences. Also, there has been attempts~\cite{koch2013approximate,harbers2015embedding,friedman2002value} to integrate values in software, e.g., the work of Friedman et al. \cite{friedman2002value} on Value Sensitive Design. But the solutions provided by these works mainly focus on early stages of software development and do not translate well into software code, thus making it hard to verify~\cite{mougouei2018operationalizing}. 

Failure to account for human values in software may lead to breaching those values and introducing biases in software~\cite{ferrario2016values,Winter2018}. Examples include falsely labeling black defendants as potential criminals by the recidivism assessment models of the US criminal justice system \cite{Bellamy2019}, higher error rate in detecting dark-skinned women compared to light-skinned men in facial recognition software \cite{Bellamy2019}, and bias against women in a popular job-recruiting software \cite{Bellamy2019}. Similar issues can be prevented if software developers are informed about the value implications of their choices and take initiatives to mitigate value breaches in software. To this end, we propose the notion of \textit{Value Programming} with three main principles: \textbf{(P1)} annotating source code and its related artifacts with respect to human values; \textbf{(P2)} inspecting source code to detect conditions that can lead to value breaches in software (i.e. \textit{Value Smells}); and \textbf{(P3)} making recommendations to mitigate value breaches.

To enable value programming in software projects, we contribute a framework referred to as \textit{AIR} (Annotation, Inspection, and Recommendation). At the heart of the framework is a catalog of commonly adopted APIs. The APIs are annotated based on their relevance to values, e.g., Android Accessibility Service can be attributed to value Benevolence). We especially focus on APIs as they are commonly used across software projects~\cite{DelaMora2018}. Moreover, it has been widely recognized that APIs are not value-agnostic; the relevance of APIs to human values such as Benevolence (via Accessibility and Transparency), self-direction (via privacy), and security has long been recognized in the literature~\cite{Pektas2019,naseri2019accessileaks,rashidi2017android,Ito2018,Olejnik2016}. 

Our proposed framework thus uses the \textit{Value-Annotated} APIs and their usage patterns~\cite{Pektas2019} to specify the relevance of the code elements (e.g., classes, functions, and data) to values and annotate them~\cite{lima2018metrics} accordingly. This allows for inspecting the source code to detect value smells (e.g., lack of Unicode support may result in breaching Universalism in a messaging system), and making recommendations, when appropriate, to mitigate those value breaches (e.g., recommending libraries for Unicode support to mitigate the risk of breaching universalism). 

\begin{figure*}[!htbp]
	\centering
	\hspace{-2cm}\includegraphics[scale=0.225,angle=0]{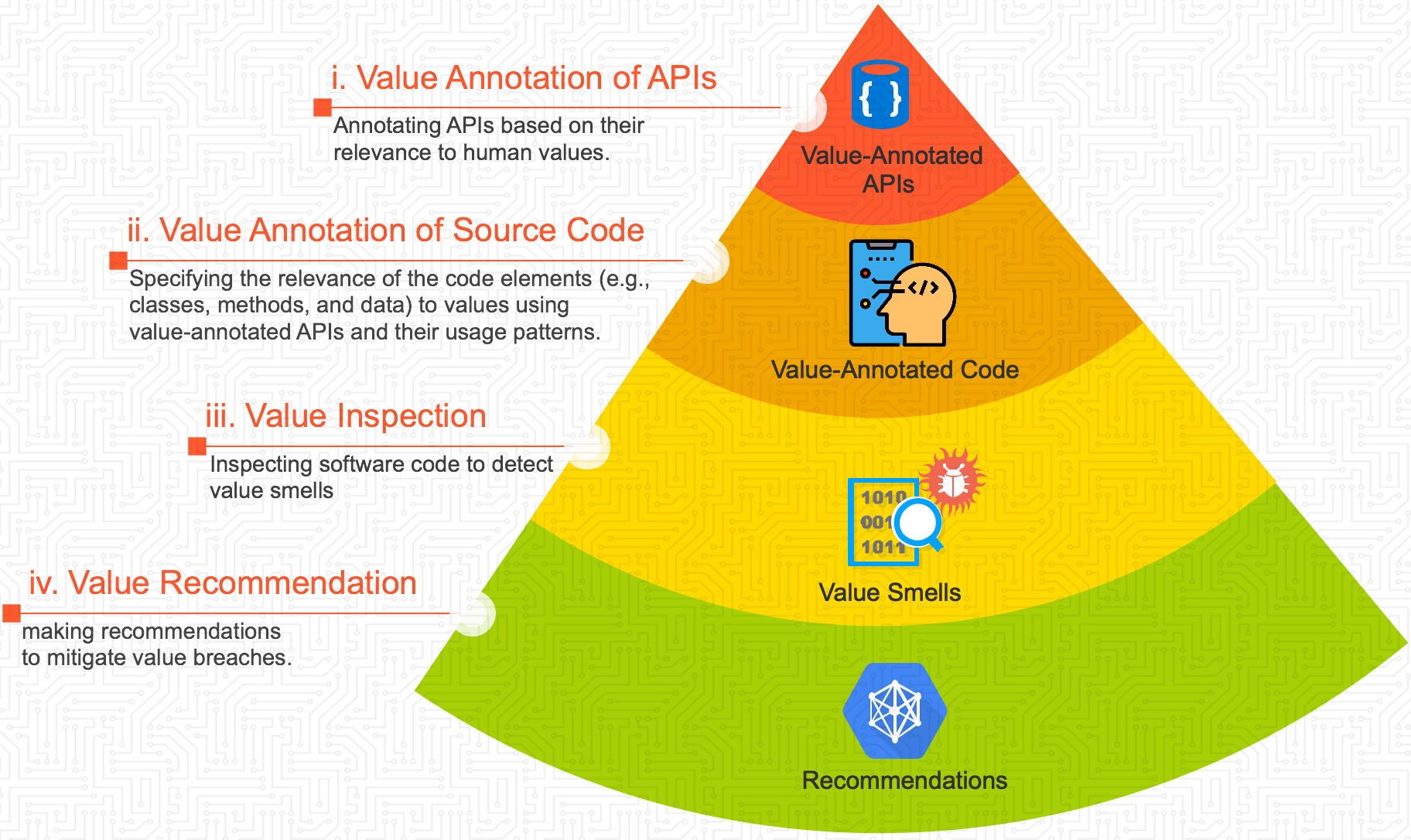}
	\captionsetup{margin=0cm}
	\caption{An overview of the AIR framework.}
	\label{fig_air}
\end{figure*}

%% file: valueprogramming.tex
\section{Value Programming}
\label{sec_valueprogramming}


In an effort to account for human values in software, we propose the notion of Value Programming, which is defined as \textit{a measurable integration of human values in software code and related artifacts}. The three principles of value programming are as follows.

\begin{itemize}[leftmargin=0.1cm]
	\item [] \textbf{(P1)} Value Annotation: specifying the relevance of code elements (e.g., classes and methods) to human values.
	
	\item [] \textbf{(P2)} Value Inspection: inspecting source code to detect conditions that lead to potential value breaches in software (value smells).
	
	\item [] \textbf{(P3)} Value Recommendation: making recommendations to address values and mitigate value breaches and biases in software. 
\end{itemize}


%% file: proposed.tex
\section{The Proposed Framework}
\label{sec_vph}

To enable value programming in software development, we have proposed a framework referred to as AIR (Annotation, Inspection, and Recommendation). The proposed framework is comprised of four major components as depicted in Figure~\ref{fig_air}: (i) annotating commonly adopted APIs based on their relevance to human values, (ii) annotating code elements (e.g., classes, methods, and data) based on the usage of value-annotated APIs, (iii) inspecting code to detect value smells (conditions that can lead to value breaches), and (iv) making recommendations to mitigate the value breaches in software. In the following, we describe the components of AIR.

\subsection{Value Annotation of APIs}

APIs are widely used in software projects. But using those APIs without considering their value implications may lead to breaching human values. Privacy leaks attributed to APIs are examples of such breaches~\cite{naseri2019accessileaks,Liu2012}. To mitigate value breaches, APIs can be explicitly annotated with their value implications. The artifacts associated with an API (e.g., API documentation) can be used to identify the relevance of that API to human values~\cite{gu2016deep}. Table~\ref{valueLog}, for instance, gives the relevance of some of the Android APIs to human values based on the documentation of those APIs\footnote{https://developer.android.com/reference}. Value annotation of APIs can be automated using machine learning techniques~\cite{gu2016deep}. Value-Annotated APIs will be used for annotating the elements of software code (e.g., classes) with respect to human values.   

Also, value-annotated APIs can help detect value dependencies~\cite{mougouei2020dependency,mougoueidependency,mougouei2017modeling,mougouei2017dependency,Mougouei2016} and conflicts in software code~\cite{naseri2019accessileaks}. This is particularly important for reconciling value conflicts (tensions) in software projects. For instance, a study by Naseri et al.~\cite{naseri2019accessileaks} showed that enabling Android Accessibility Service\footnote{https://developer.android.com/reference/android/accessibilityservice/package-summary} leaves 72\% of the top finance and 80\% of the top social media apps vulnerable to eavesdropping attacks and leaking sensitive information. Hence, as in Table~\ref{valueLog}, Android Accessibility Service values Benevolence (via Helpfulness) and Universalism (via Equality) while having a negative impact on Security and Self Direction (via privacy leaks). These can be captured by annotating the API \footnote{https://developer.android.com/studio/write/annotations}.  

%
\begin{table*}[!htbp]
	\caption{Human values and their definitions based on Schwartz theory of basic values~\cite{lee2019testing}.}
	\label{values}
	\centering
	\input{values}
\end{table*}
\begin{table*}[!htbp]
	\caption{The relevance of android APIs to values: relations can be positive ($+$), negative ($-$), unknown ($\pm$), or non-relevant (empty cell).}
	\label{valueLog}
	\centering
	\input{valuelog}
\end{table*}


\subsection{Value Annotation of Code}

Value annotation of code allows for analyzing and improving software programs with respect to human values (e.g., Benevolence). For instance, Java's mechanism for annotation processing~\cite{7321546} can be used to automatically analyze values in code elements. This, however, does not affect the execution of a program per se. There are several advantages to value annotation of code as follows.

\begin{itemize}[leftmargin=0.3cm]
	\item \textit{Interoperability.} Annotation is widely supported across different programming languages (e.g., Java and Python); we can leverage from this to embed value programming in software development.
	\item \textit{Extensibility.} Value annotations can be extended/customized to cope with different interpretations of human values~\cite{mougouei2018operationalizing}. 
	\item \textit{Static Analysis.} Code annotation has been used for compile-time checking of privacy~\cite{zimmeck2016automated}. This can be extended to other values. 
	\item \textit{Automation.} Value annotations can be used to generate code that adheres to the values of the users.  
	\item \textit{Standardization.} Value annotation allows for uniformly describing the relevance of code elements to values across different platforms. 
\end{itemize}

Value annotation of software code can be achieved by enriching the code elements with metadata that specifies the relevance of those elements to human values (Table~\ref{values}). This can be performed manually or automatically. The proposed framework (AIR) focuses on automated annotation of code based on the usage of value-annotated APIs, as explained earlier. The process starts with identifying value-annotated APIs in software code. Then the elements of the code that interact with those APIs will be annotated accordingly. In other words, if an API relates to a value, so do the code elements that interact with that API. 

\begin{figure}[!htbp]
	\centering
	\includegraphics[scale=0.24,angle=0]{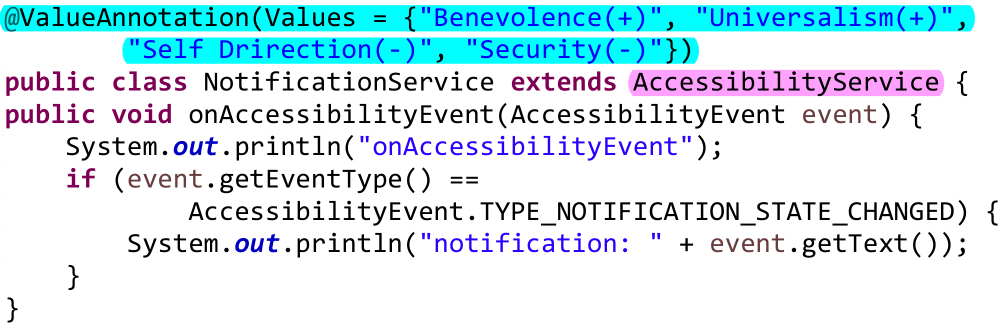}
	\captionsetup{margin=0cm}
	\vspace{-0.5cm}
	\caption{Annotation of a class based on the values related to Android Accessibility Service (Table~\ref{valueLog}). Code is from Stack Overflow.}
	\label{fig_example1}
\end{figure}

For instance, classes and methods that use different features of the Android Accessibility Service can be annotated as relevant to Benevolence, Universalism, Self Direction, and Security (Table~\ref{valueLog}). An example is given in Figure~\ref{fig_example1}: Java annotation package (``java.lang.annotation'') is used to create a customized annotation interface (@ValueAnnotation) and value annotate class ``NotificationService'', which extends Android Accessibility Service. 

\subsection{Value Inspection}

Value annotation of source code allows for automated analysis of values. In this regard, value annotation processors can be used to analyze the code elements that relate to values. For instance, class ``NotificationService'' in Figure~\ref{fig_example1} extends Android Accessibility Service, thus annotated as positively relevant to Benevolence and Universalism, and negatively related to Self Direction and Security (Table~\ref{valueLog}). Vulnerabilities/faults in ``NotificationService', therefore, impact these values. Hence, inspection can be carried out to identify the conditions that can breach the values associated with ``NotificationService''. Such conditions are referred to as value smells. Hence, Value Inspection aims to identify value smells in software code.

\subsection{Value Recommendation}

The Value Recommendation component of AIR aims to mitigate conditions that may lead to breaching human values and introducing biases in software. These conditions (value smells) are detected by the Value Inspection component, as discussed earlier, and recommendations will be made as to how to mitigate them. Such recommendations can be high level or low level (e.g., direct fixes). For value item Privacy, Nseri et al.~\cite{naseri2019accessileaks} proposed a tool named AcFix that automatically fixes certain privacy leaks associated with Android Accessibility Service in mobile apps. These fixes are performed based on recommendations such as ``Apps can prevent information leaks through accessibility services by setting a single flag for sensitive inputs.'' AcFix parses Java source code and builds an Abstract Syntax Tree, which is exploited to analyze the app source code and to detect the sensitive user input fields, which can be eavesdropped through the accessibility service. Similar tools can be devised for recommending fixes to mitigate other value smells in software code. It is important, however, that conflicts among human values are taken into account when making recommendations. In the case of AcFix~\cite{naseri2019accessileaks}, for instance, the authors identified that ``Disabling accessibility services for a field improves security but reduces accessibility.''

%% file: values.tex
\Huge\resizebox {0.98\textwidth }{!}{
\begin{tabular}{|l|l|l|l|}
	\toprule
	\hline
	\rowcolor[HTML]{000000} 
	\multicolumn{1}{|l|}{\cellcolor[HTML]{000000}{\color[HTML]{FFFFFF} \textbf{ID}}}&
	\multicolumn{1}{|l|}{\cellcolor[HTML]{000000}{\color[HTML]{FFFFFF} \textbf{Value}}} & \multicolumn{1}{l|}{\cellcolor[HTML]{000000}{\color[HTML]{FFFFFF} \textbf{Definition}}} & \multicolumn{1}{l|}{\cellcolor[HTML]{000000}{\color[HTML]{FFFFFF} \textbf{Descriptive Value Items}}} \\ \hline
	
	
	\multirow{2}{*}{$V_1$}
	& \multirow{2}{*}{Self Direction} 
	&  thought: freedom to cultivate one's own ideas and abilities.        
	&  creativity/imagination; curious/interested \\ \cline{3-4} &
	&  action: freedom to determine one's own actions.       
	&  choosing own goals/own purposes; independent/self-reliant; privacy \\ \hline
	
	$V_2$
	&\multicolumn{1}{|l|}{\begin{tabular}[c]{@{}l@{}}Stimulation\end{tabular}} 
	& \multicolumn{1}{l|}{excitement, novelty, and change.} 
	& \multicolumn{1}{l|}{\begin{tabular}[c]{@{}l@{}} an exciting life/stimulating experiences; a varied life; daring/seeking adventure\end{tabular}} \\ \hline
		
	$V_3$
	& \multicolumn{1}{|l|}{\begin{tabular}[c]{@{}l@{}}Hedonism\end{tabular}} 
	& \multicolumn{1}{l|}{pleasure and sensuous gratification} 
	& \multicolumn{1}{l|}{\begin{tabular}[c]{@{}l@{}} pleasure\end{tabular}} \\ \hline
	
	$V_4$
	& \multicolumn{1}{|l|}{\begin{tabular}[c]{@{}l@{}}Achievement\end{tabular}} 
	& \multicolumn{1}{l|}{success according to social standards} 	
	& \multicolumn{1}{l|}{\begin{tabular}[c]{@{}l@{}}successful/achieving goals; ambitious/aspiring; capable/ competent;\\ influential/having an impact on people and events\end{tabular}} \\ \hline
	
	\multirow{2}{*}{$V_5$}
	& \multirow{2}{*}{Power} 
	&  dominance: exercising control over people.     
	&  social power/control over others; authority/right to command \\ \cline{3-4} & 
	&  resources: control of material and social resources.   
	&  wealth/material possessions \\ \hline

    $V_6$
    & \multicolumn{1}{|l|}{\begin{tabular}[c]{@{}l@{}}Face\end{tabular}} 
	& \multicolumn{1}{l|}{maintaining one's public image and avoiding humiliation.} 
	& \multicolumn{1}{l|}{\begin{tabular}[c]{@{}l@{}}social recognition/respect; preserving public image/maintaining face \end{tabular}} \\ \hline
	
	\multirow{2}{*}{$V_7$}
	& \multirow{2}{*}{Security} 
	&  personal: safety in one's immediate environment.
	&  \begin{tabular}[c]{@{}l@{}}sense of belonging/feeling others care about me; healthy/ not sick; \\reciprocating favors/avoiding indebtedness; clean/neat, tidy, \\ family security/safety for loved ones* \end{tabular}\\ \cline{3-4} & 
	&  social: safety and stability in the wider society 
	&  national security/nation safe from enemies; social order/societal stability \\ \hline
	
	$V_8$
	& \multicolumn{1}{|l|}{Tradition} 
	& \multicolumn{1}{l|}{\begin{tabular}[c]{@{}l@{}}maintaining and preserving cultural, family or religious traditions\end{tabular}} 
	& \multicolumn{1}{l|}{\begin{tabular}[c]{@{}l@{}} respect tradition/preserve customs; devout/hold religious faith\end{tabular}} \\ \hline
	
	\multirow{2}{*}{$V_9$}
	& \multirow{2}{*}{Conformity} 
	&  rules: compliance with rules, laws, and formal obligations.
	&  self-discipline/resist temptation; obedient/meet obligations \\ \cline{3-4} & 
	&  interpersonal: avoidance of upsetting or harming other people   
	&  politeness/courtesy; honor parents/show respect \\ \hline
	
	$V_{10}$
	& \multicolumn{1}{|l|}{\begin{tabular}[c]{@{}l@{}}Humility\end{tabular}} 
	& \multicolumn{1}{l|}{recognizing one's insignificance in the larger scheme of things.} 
	& \multicolumn{1}{l|}{\begin{tabular}[c]{@{}l@{}}humble/modest, self-effacing; accepting my portion/ \\submitting to life's circumstances \end{tabular}} \\ \hline
	
	\multirow{2}{*}{$V_{11}$}
	& \multirow{2}{*}{Benevolence} 
	&  caring: devotion to the welfare of in-group members.
	&  \begin{tabular}[c]{@{}l@{}}helpful/working for others welfare; honest/genuine;\\ forgiving/willing to pardon; \\family security/safety for loved ones* \end{tabular} \\ \cline{3-4} & 
	&  dependability: being a reliable and trustworthy member of the in-group
	&  responsible/dependable; loyal/faithful to friends \\ \hline
	
	\multirow{2}{*}{$V_{12}$}
	& \multirow{2}{*}{Universalism} 
	&  nature: preservation of the natural environment.
	&  protect the environment; unity with nature; world beauty \\ \cline{3-4} & 
	&  concern: commitment to equality, justice and protection for all people
	&  equality for all; social justice; world at peace \\  \cline{3-4} & 
	&  tolerance: acceptance and understanding of those who are different from oneself
	&  broadminded/tolerant; wisdom/mature understanding \\ \hline

\end{tabular}	
}

%% file: valuelog.tex
\large
\resizebox {0.975\textwidth }{!}{
\begin{tabular}{|l|l|l|l|l|l|l|l|l|l|l|l|l|l|}
	\toprule
	\hline
	\rowcolor[HTML]{333333} 
	{\color[HTML]{FFFFFF} \textbf{API}} & {\color[HTML]{FFFFFF} \textbf{Description}} & {\color[HTML]{FFFFFF} \textbf{$V_1$}} & {\color[HTML]{FFFFFF} \textbf{$V_2$}} & {\color[HTML]{FFFFFF} \textbf{$V_3$}} & {\color[HTML]{FFFFFF} \textbf{$V_4$}} & {\color[HTML]{FFFFFF} \textbf{$V_5$}} & {\color[HTML]{FFFFFF} \textbf{$V_6$}} & {\color[HTML]{FFFFFF} \textbf{$V_7$}} & {\color[HTML]{FFFFFF} \textbf{$V_8$}} & {\color[HTML]{FFFFFF} \textbf{$V_9$}} & {\color[HTML]{FFFFFF} \textbf{$V_{10}$}} & {\color[HTML]{FFFFFF} \textbf{$V_{11}$}} & {\color[HTML]{FFFFFF} \textbf{$V_{12}$}} \\ \hline
	
	android.accessibilityservice 
	& \multicolumn{1}{l|}{\begin{tabular}[c]{@{}l@{}}... used to assist users with disabilities in \\using Android devices and apps ... . \end{tabular}}                    
	& $-$ & & & 
	& & & $-$ & 
	& & & $+$ & $+$ \\ \hline
	
    android.animation 
	& \multicolumn{1}{l|}{\begin{tabular}[c]{@{}l@{}}... provide functionality for the property animation system, \\which allows you to animate object properties of any type. \end{tabular}}                    
	& & & $\pm$ & 
	& & & & 
	& & & &\\ \hline
	
	android.app.admin 
	& \multicolumn{1}{l|}{\begin{tabular}[c]{@{}l@{}} ... provides device administration features at the system level, \\ allowing you to create security-aware applications that are useful in enterprise settings, \\ in which IT professionals require rich control over employee devices.
    \end{tabular}}                    
	& & &  & 
	& $+$ & & $+$ & 
	& & & &\\ \hline
	
	android.app.role 
	& \multicolumn{1}{l|}{\begin{tabular}[c]{@{}l@{}} ... provides information about and manages roles.
	\end{tabular}}                    
	&  $+$ & &  & 
	& $+$ & & $+$ & 
	& & & &\\ \hline
	
	android.icu.lang 
	& \multicolumn{1}{l|}{\begin{tabular}[c]{@{}l@{}} international language support.
	\end{tabular}}                    
	&  & &  & 
	& &  & & 
	& & & $+$& $+$\\ \hline
	
	android.icu.media 
	& \multicolumn{1}{l|}{\begin{tabular}[c]{@{}l@{}} ... provides classes that manage various media interfaces in audio and video.\\
	The Media APIs are used to play and, in some cases, record media files. \\ Other special classes in the package offer the ability to detect the faces ..., \\ control audio routing (to the device or a headset) and control alerts such as \\ ringtones and phone vibrations (AudioManager) ... .
	\end{tabular}}                    
	& $\pm$  & $\pm$ & $\pm$ & 
	& &  & & 
	& & & & $\pm$\\ \hline
	
	android.mtp 
	& \multicolumn{1}{l|}{\begin{tabular}[c]{@{}l@{}}... let you interact directly with connected cameras and other devices ... . \\Your application can receive notifications when devices are attached and removed, \\ manage files and storage on those devices, and transfer files and metadata from the devices.
	\end{tabular}}                    
	& $\pm$ &  &  & 
	& &  &  & 
	& & &  & \\ \hline

	android.nfc 
	& \multicolumn{1}{l|}{\begin{tabular}[c]{@{}l@{}}provides access to Near Field Communication (NFC) functionality\\, allowing applications to read NDEF message in NFC tags. \\ A "tag" may actually be another device that appears as a tag.
	\end{tabular}}                    
	& $\pm$ &  &  & 
	& &  & $\pm$  & 
	&$+$ & &  & \\ \hline
	
    android.security 
	& \multicolumn{1}{l|}{\begin{tabular}[c]{@{}l@{}}provides access to a few facilities of the Android security subsystems.
	\end{tabular}}                    
	&  &  &  & 
	& &  & $+$  & 
	& & &  & \\ \hline 
%
%
%
\end{tabular}
}

%% file: conclusion.tex
\section{Summary}
\label{sec_conclusion}

This paper proposed the notion of Value Programming, defined as \textit{a measurable integration of human values in software code and related artifacts}. To enable value programming in software development, we proposed a framework comprising of four major components: (i) annotating commonly adopted APIs based on their relevance to human values; (ii) annotating code elements (e.g., classes and methods) based on their interactions with the value-annotated APIs; (iii) inspecting code to detect value smells (conditions that can lead to value breaches); and (iv) making recommendations to mitigate the value breaches in software. We are currently implementing a prototype of AIR to value-annotate real-world software code, thus allowing for the evaluation of AIR. 

